\documentclass[11pt]{article}
\usepackage{geometry}
\usepackage[nosort]{cite}
\usepackage{amsmath,amssymb}

\makeatletter \@addtoreset{equation}{section} \makeatother

\newcommand{\fft}[2]{{\frac{#1}{#2}}}
\newcommand{\ft}[2]{{\textstyle\frac{#1}{#2}}}

\begin{document}
\begin{flushright}
MCTP-08-52
\end{flushright}

\vspace{10pt}
\begin{center}

{\Large\bf Higher derivative corrections to $R$-charged AdS$_5$ black holes
and field redefinitions}

\vspace{20pt}

James T. Liu and Phillip Szepietowski

\vspace{20pt}

{\it Michigan Center for Theoretical Physics\\
Randall Laboratory of Physics, The University of Michigan\\
Ann Arbor, MI 48109--1040}

\vspace{40pt}

\underline{ABSTRACT}
\end{center}

We consider four-derivative corrections to the bosonic sector of
five-dimensional $\mathcal N=2$ gauged supergravity.  Since this theory
includes the $\mathcal N=2$ graviphoton, we consider both curvature and
graviphoton field-strength terms that show up at the four-derivative level.
We construct, to linear order, the higher-derivative corrections to the
non-rotating $R$-charged AdS$_5$ black hole and demonstrate how this
solution transforms under field redefinitions.

\newpage

\section{Introduction}

Higher derivative corrections to the Einstein-Hilbert action have
received much notice in recent years, as such terms naturally show
up in the $\alpha'$ expansion of effective actions derived from string
theory.  In general, the first non-trivial terms arise at the four
derivative level, corresponding to curvature-squared corrections to
classical Einstein theory of the form
\begin{equation}
e^{-1}\delta\mathcal L=\alpha_1R^2 + \alpha_2R_{\mu\nu}R^{\mu\nu}
+ \alpha_3R_{\mu\nu\rho\sigma}R^{\mu\nu\rho\sigma},
\label{eq:la1a2a3}
\end{equation}
where the coefficients $\alpha_1$, $\alpha_2$ and $\alpha_3$ are
determined by the underlying theory.  It was suggested in
\cite{Zwiebach:1985uq} that the natural form of such terms would be given
by the Gauss-Bonnet combination
\begin{equation}
e^{-1}\delta\mathcal L_{\rm GB}=\alpha(R^2-4R_{\mu\nu}R^{\mu\nu}
+R_{\mu\nu\rho\sigma}R^{\mu\nu\rho\sigma}),
\label{eq:lgb}
\end{equation}
as this is the unique combination that avoids
introducing ghosts in the effective theory.  It was subsequently
argued, however, that in the absence of an off-shell formulation such
as string field theory, the $\alpha_1$ and $\alpha_2$ coefficients are
physically indeterminate as they may be eliminated by an on-shell field
redefinition of the form $g_{\mu\nu}\to g_{\mu\nu}+aRg_{\mu\nu}+bR_{\mu\nu}$.
In this sense, only the Riemann-squared term parameterized by $\alpha_3$
carries physical information from the underlying string theory.

The form of the higher derivative corrections are further constrained by
supersymmetry.  Explicit computations for the uncompactified closed
superstring indicate that the first corrections enter at the $R^4$
order \cite{Gross:1986iv,Grisaru:1986vi,Freeman:1986zh}.  This
is a feature of maximal supersymmetry, as curvature-squared terms are
present in, for example, the uncompactified heterotic theory
\cite{Gross:1986mw,Metsaev:1987zx}.  An alternate route to obtaining
supersymmetric higher derivative corrections is to make use of supersymmetry
itself to construct higher derivative invariants that may show up in the
action.  This was applied in the heterotic supergravity by supersymmetrizing
the Lorentz Chern-Simons form responsible for the modified Bianchi
identity $dH=\alpha'\mbox{Tr\,}(F\wedge F-R\wedge R)$ \cite{Bergshoeff:1989de};
the result agrees with the explicit calculations, once field redefinitions
are properly taken into account \cite{Chemissany:2007he}.  More recently,
the supersymmetric completion of the $A\wedge \mbox{Tr\,} R\wedge R$ term
in five-dimensional $\mathcal N=2$ supergravity (coupled to a number of
vector multiplets) was obtained in \cite{Hanaki:2006pj}.  This result has
led to new progress in the study of black hole entropy and precision
microstate counting in five dimensions (see {\it e.g.}~\cite{Castro:2008ne}
and references therein).

The supersymmetric four-derivative terms given in \cite{Hanaki:2006pj}
were obtained using conformal supergravity methods.  Thus it should be
no surprise that they involve the square of the five-dimensional
Weyl tensor \cite{Hanaki:2006pj}
\begin{eqnarray}
e^{-1}\delta\mathcal L_{\rm sugra}&=&\ft{c_I}{24}[\ft18M^IC_{\mu\nu\rho\sigma}
C^{\mu\nu\rho\sigma}+\cdots]\nonumber\\
&=&\ft{c_I}{24}[\ft18M^I(\ft16R^2-\ft43R_{\mu\nu}
R^{\mu\nu}+R_{\mu\nu\rho\sigma}R^{\mu\nu\rho\sigma})+\cdots],
\label{eq:hanaki}
\end{eqnarray}
as opposed to the Gauss-Bonnet combination, (\ref{eq:lgb}).  In principle,
an appropriate field redefinition may be performed to bring this into the
Gauss-Bonnet form.  However, this is usually not done, as it would obscure
the overall supersymmetric structure of the theory.  Thus in practice two
somewhat complimentary approaches have been taken to investigating the
curvature-squared corrections to the Einstein-Hilbert action.  The first,
which applies whether the underlying theory is supersymmetric or not, is
to use a parameterized action of the form (\ref{eq:la1a2a3}), with special
emphasis on the Gauss-Bonnet combination.  The second is to focus
directly on supergravity theory, and hence to use explicitly supersymmetric
higher-derivative actions of the form (\ref{eq:hanaki}).  In principle,
these two approaches are related by appropriate field redefinitions.  However,
in practice this is complicated by the fact that additional matter
fields ({\it e.g.}~$\mathcal N=2$ vector multiplets) as well as auxiliary
fields may be present, thus making any field redefinition highly non-trivial.

In this letter, we investigate and clarify some of the issues surrounding
field redefinitions in the presence of additional fields.  In particular,
we take the bosonic sector of five-dimensional $\mathcal N=2$ gauged
supergravity and extend it with four-derivative terms built from the
Riemann tensor $R_{\mu\nu\rho\sigma}$ as well as the graviphoton
field-strength tensor $F_{\mu\nu}$.  Although we introduce eight such
terms, we demonstrate that only four independent combinations remain
physical once field redefinitions are taken into account.  To be explicit,
we construct the higher-derivative corrections to the spherically symmetric
$R$-charged AdS$_5$ black holes of \cite{Behrndt:1998ns,Behrndt:1998jd},
working to linear order in the higher-derivative terms, and then investigate
the effect of field redefinitions on these black hole solutions.

To some extent, our solutions generalize the Gauss-Bonnet black holes
originally constructed in \cite{Boulware:1985wk,Wheeler:1985nh} and
extended to Einstein-Maxwell theory in \cite{Wiltshire:1985us} and,
with the inclusion of Born-Infeld terms, in \cite{Wiltshire:1988uq}.
One advantage that the Gauss-Bonnet combination has over the generic
form of (\ref{eq:la1a2a3}) is that it leaves the graviton propagator
unmodified, and also yields a modified Einstein equation involving at
most second derivatives of the metric.  With an appropriate metric ansatz,
the resulting Gauss-Bonnet black holes are then obtained by solving a
simple quadratic equation.  Furthermore, this feature of the Gauss-Bonnet
term leads to a good boundary variation and natural generalization of
the Gibbons-Hawking surface term \cite{Myers:1987yn}.  This is a primary
reason behind the popularity of applying Gauss-Bonnet (and more generally
Lovelock) extensions to braneworld physics (see
{\it e.g.}~\cite{Charmousis:2008kc}).

Our interest in studying the higher order corrections to $R$-charged
AdS$_5$ black holes is also motivated by our desire to explore finite
't~Hooft coupling corrections in AdS/CFT.  Using the relation $\alpha'
=L^2/\sqrt\lambda$, we see that each additional factor of
$\alpha' R_{\mu\nu\rho\sigma}$ in the string effective action gives rise
to a $1/\sqrt\lambda$ factor in the strong coupling expansion of the
dual gauge theory.  Since supersymmetry ensures that the leading correction
terms in IIB theory are of order $\alpha'^3$, this indicates that the
$\mathcal N=4$ super-Yang Mills theory dual to AdS$_5\times S^5$ will first
receive such corrections at the $\lambda^{-3/2}$ order.  The effect of
these finite 't~Hooft coupling corrections on both the thermodynamics
\cite{Gubser:1998nz,Pawelczyk:1998pb} and hydrodynamics
\cite{Buchel:2003tz,Buchel:2004di,Benincasa:2005qc,Buchel:2008ac,Buchel:2008wy,Buchel:2008sh}
of the $\mathcal N=4$ plasma have received much attention in the context of
extrapolations between the strong and weak coupling limits of the
$\mathcal N=4$ theory.

In principle, it would be greatly desirable to extend the finite coupling
analysis to $\mathcal N=1$ gauge theories dual to AdS$_5\times Y^5$ where
$Y^5$ is Sasaki-Einstein.  This is of particular interest in resolving
conjectures on the nature of the shear viscosity bound $\eta/s$
\cite{Policastro:2001yc,Kovtun:2003wp,Buchel:2003tz,Kovtun:2004de,Kats:2007mq,Brigante:2007nu,Brigante:2008gz}.
One difficulty in doing so, however, lies in the fact that the higher
derivative corrections involving the Ramond-Ramond five-form have not yet
been fully explored (but see \cite{Paulos:2008tn}).  While it may be
argued that these terms will not contribute in the maximally supersymmetric
case, there is no reason to expect this to continue to hold for the reduced
supersymmetric backgrounds dual to $\mathcal N=1$ super-Yang Mills.  For
this reason, recent investigations of the shear viscosity
\cite{Kats:2007mq,Brigante:2007nu,Brigante:2008gz}
(and drag force \cite{Fadafan:2008gb,VazquezPoritz:2008nw})
have assumed a
parameterized set of curvature-squared corrections of the form indicated
above in (\ref{eq:la1a2a3}).  Our present construction of higher-derivative
corrected $R$-charged black holes allows for a generalization of the
finite coupling shear viscosity calculation to backgrounds dual to turning
on a chemical potential \cite{Benincasa:2006fu}.

We start with the two-derivative bosonic action of $\mathcal N=2$ gauged
supergravity and in Section~2 we introduce a parameterized set of four
derivative terms involving both curvature and graviphoton field strengths.
Then, in Section~3, we obtain the linearized corrections to the spherically
symmetric $R$-charged AdS$_5$ black holes.  As one of the aims of this letter
is to clarify the use of field redefinitions, we take a closer look at this
in Section~4.  Finally, we conclude with a discussion of our results in
Section~5.

\section{The higher-derivative theory}

Our starting point is the bosonic sector of pure $\mathcal N=2$ gauged
supergravity in five dimensions, with Lagrangian given by
\begin{equation}
e^{-1}{\mathcal L}_0 = R - \ft14F_{\mu\nu}F^{\mu\nu}+ 12g^2 +\ft1{12\sqrt3}
\epsilon^{\mu\nu\rho\sigma\lambda}F_{\mu\nu}F_{\rho\sigma}A_{\lambda}.
\label{eq:lag0}
\end{equation}
Although the Chern-Simons term is important from a supergravity point of view,
it will not play any role in the electrically charged solutions that are
investigated below.

In general, higher-derivative corrections to $\mathcal L_0$ may be expanded
in the number of derivatives.  We are mainly interested in the first non-trivial
corrections, which arise at the four-derivative level.  In a pure gravity
theory, this would correspond to the addition of $R^2$ terms to the Lagrangian. However, for the Einstein-Maxwell system, we may also consider higher-order
terms in the Maxwell field, such as $F^4$ and $RF^2$ terms.  We thus introduce
the higher-derivative Lagrangian
\begin{equation}
{\cal L} = {\cal L}_0 + {\cal L}_{R^2} + {\cal L}_{F^4} + {\cal L}_{RF^2},
\label{eq:lhd}
\end{equation}
where $ {\cal L}_0 $ is given in (\ref{eq:lag0}), while the additional
terms are
\begin{eqnarray}
e^{-1}{\cal L}_{R^2} &=& \alpha_1R^2 + \alpha_2R_{\mu\nu}R^{\mu\nu}
+ \alpha_3R^{\mu\nu\rho\sigma}R_{\mu\nu\rho\sigma},\nonumber\\
e^{-1}{\cal L}_{F^4} &=& \beta_1(F_{\mu\nu}F^{\mu\nu})^2 +
\beta_2F^{\mu}{}_{\nu}F^{\nu}{}_{\rho}F^{\rho}{}_{\sigma}F^{\sigma}{}_{\mu},
\nonumber\\
e^{-1}{\cal L}_{RF^2} &=& \gamma_1RF_{\mu\nu}F^{\mu\nu} +
\gamma_2R_{\mu\nu}F^{\mu\rho}F_{\rho}{}^{\nu} +
\gamma_3R^{\mu\nu\rho\sigma}F_{\mu\nu}F_{\rho\sigma}.
\label{eq:hdlag}
\end{eqnarray}
Note that we have not considered terms such as $F_{\mu\nu}\Box F^{\mu\nu}$
that would in principle enter at the same order.  Although we are not
complete in this regard, the terms that enter in $\mathcal L_{F^4}$ are
nevertheless sufficient for capturing the expansion of the Born-Infeld action.

\subsection{Equations of Motion}

Both the Maxwell and Einstein equations pick up corrections from the
higher-derivative terms in (\ref{eq:lhd}).  The modified Maxwell equation
is straightforward
\begin{eqnarray}
\nabla_\mu F^{\mu\nu}+\ft1{4\sqrt3}\epsilon^{\nu\rho\lambda\sigma\delta}
F_{\rho\lambda}F_{\sigma\delta}&=&\nabla_\mu\bigl(8\beta_1F^2 F^{\mu\nu}
-8\beta_2F^{\mu\lambda}F_{\lambda\sigma}F^{\sigma\nu}\nonumber\\
&&\qquad+4\gamma_1 RF^{\mu\nu}+4\gamma_2(R^{[\mu}{}_\lambda F^{\nu]\lambda})
+4\gamma_3R^{\mu\nu\lambda\sigma}F_{\lambda\sigma}\bigr).\qquad
\label{eq:max}
\end{eqnarray}
The Einstein equation is somewhat cumbersome, but can be expressed in Ricci
form as
\begin{eqnarray}
&&R_{\mu\nu}+4g^2g_{\mu\nu}-\ft12F_{\mu\lambda}F_\nu{}^\lambda
+\ft1{12}g_{\mu\nu}F^2=\nonumber\\
&&\kern4em(2\alpha_1+\alpha_2+2\alpha_3)\nabla_\mu\nabla_\nu R
-(\alpha_2+4\alpha_3)\Box R_{\mu\nu}\nonumber\\
&&\kern4em-2\alpha_1RR_{\mu\nu}+4\alpha_3R_{\mu\lambda}R_\nu{}^\lambda
-2(\alpha_2+2\alpha_3)R_{\mu\lambda\nu\sigma}R^{\lambda\sigma}
-2\alpha_3R_{\mu\rho\lambda\sigma}R_\nu{}^{\rho\lambda\sigma}\nonumber\\
&&\kern4em+\ft13g_{\mu\nu}[(2\alpha_1+\alpha_2+2\alpha_3)\Box R+\alpha_1R^2
+\alpha_2R_{\lambda\sigma}^2+\alpha_3R_{\rho\lambda\sigma\delta}^2]\nonumber\\
&&\kern4em-4\beta_1F^2F_{\mu\lambda}F_\nu{}^\lambda
-4\beta_2F_{\mu\rho}F^{\rho\lambda}F_{\lambda\sigma}F^\sigma{}_\nu
+g_{\mu\nu}[\beta_1(F^2)^2+\beta_2F^4]\nonumber\\
&&\kern4em+\gamma_1(\nabla_\mu\nabla_\nu F^2-R_{\mu\nu}F^2
-2RF_{\mu\lambda}F_\nu{}^\lambda)\nonumber\\
&&\kern4em+\gamma_2(-\nabla_\lambda\nabla_{(\mu}F_{\nu)\rho}F^{\lambda\rho}
+\ft12\Box F_{\mu\lambda}F_\nu{}^\lambda+2R_{(\mu}{}^\lambda F_{\nu)}{}^\rho F_{\lambda\rho}+R_{\lambda\sigma}F_\mu{}^\lambda F_\nu{}^\sigma)\nonumber\\
&&\kern4em-\gamma_3(2\nabla^\lambda\nabla^\sigma F_{\mu\lambda}F_{\nu\sigma}+3R_{\mu\rho\lambda\sigma}F_\nu{}^\rho F^{\lambda\sigma})\nonumber\\
&&\kern4em+\ft13g_{\mu\nu}[(\gamma_1-\ft12\gamma_2)\Box F^2+2\gamma_3\nabla_\lambda\nabla_\sigma F^{\lambda\rho}F^\sigma{}_\rho\nonumber\\
&&\kern7em+2\gamma_1RF^2-2\gamma_2R_{\lambda\sigma}
F^{\lambda\rho}F^\sigma{}_\rho
+2\gamma_3R^{\rho\lambda\sigma\delta}F_{\rho\lambda}F_{\sigma\delta}].
\label{eq:eins}
\end{eqnarray}
Since we are mainly interested in obtaining corrections {\it linear} in the
parameters ($\alpha_1$, $\alpha_2$, $\alpha_3$, $\beta_1$, $\beta_2$,
$\gamma_1$, $\gamma_2$, $\gamma_3)$ of the higher derivative terms, we may
substitute the lowest order equations of motion, given by the left-hand-sides
of (\ref{eq:max}) and (\ref{eq:eins}) into the right-hand-side of
(\ref{eq:eins}) to obtain a slightly simpler form of the
Einstein equation
\begin{eqnarray}
&&R_{\mu\nu}+4g^2g_{\mu\nu}-\ft12F_{\mu\lambda}F_\nu{}^\lambda
+\ft1{12}g_{\mu\nu}F^2=\nonumber\\
&&\kern4em4g^2(5\alpha_1+\alpha_2-2\alpha_3+10\gamma_1-2\gamma_2)F_{\mu\lambda}
F_\nu{}^\lambda\nonumber\\
&&\kern4em-2\alpha_3R_{\mu\rho\lambda\sigma}R_\nu{}^{\rho\lambda\sigma}
-(\alpha_2+2\alpha_3-\gamma_2)R_{\mu\lambda\nu\sigma}F^{\lambda\rho}
F^\sigma{}_\rho-3\gamma_3R_{(\mu}{}^{\rho\lambda\sigma}F_{\nu)\rho}
F_{\lambda\sigma}\nonumber\\
&&\kern4em+\ft1{12}(2\alpha_1+\alpha_2+2\alpha_3+12\gamma_1-3\gamma_2)
\nabla_\mu\nabla_\nu F^2-\ft12(\alpha_2+4\alpha_3-\gamma_2)\Box F_{\mu\lambda}
F_\nu{}^\lambda\nonumber\\
&&\kern4em-2\gamma_3\nabla^\lambda\nabla^\sigma F_{\mu\lambda}F_{\nu\sigma}
-\ft1{12}(\alpha_1-\alpha_2+2\alpha_3+48\beta_1+8\gamma_1+2\gamma_2)
F^2F_{\mu\lambda}F_\nu{}^\lambda\nonumber\\
&&\kern4em+(\alpha_3-4\beta_2+\gamma_2)
F_{\mu\rho}F^{\rho\lambda}F_{\lambda\sigma}F^\sigma{}_\nu\nonumber\\
&&\kern4em+\ft13g_{\mu\nu}[-16g^4(5\alpha_1+\alpha_2)
-\ft23g^2(17\alpha_1+7\alpha_2+42\gamma_1-12\gamma_2)F^2\nonumber\\
&&\kern7em+\ft16(\alpha_1+2\alpha_2+7\alpha_3+6\gamma_1-3\gamma_2+3\gamma_3)
\Box F^2\nonumber\\
&&\kern7em+\ft1{144}(7\alpha_1-13\alpha_2+432\beta_1+60\gamma_1
+24\gamma_2)(F^2)^2\nonumber\\
&&\kern7em+\ft14(\alpha_2+12\beta_2-4\gamma_2)F^4
+\alpha_3R_{\rho\lambda\sigma\delta}^2+2\gamma_3R_{\rho\lambda\sigma\delta}
F^{\rho\lambda}F^{\sigma\delta}]\nonumber\\
&&\kern4em+\cdots.
\label{eq:einslin}
\end{eqnarray}
This is valid to first order in the four-derivative corrections.

Numerous previous studies higher-derivative corrections in five dimensions
have concentrated on the purely gravitational sector of the theory.  In this
case, the first order Einstein equation simplifies to
\begin{equation}
R_{\mu\nu}+4g^2g_{\mu\nu}=-2\alpha_3R_{\mu\rho\lambda\sigma}
R_\nu{}^{\rho\lambda\sigma}
+\ft13g_{\mu\nu}[-16g^4(5\alpha_1+\alpha_2)
+\alpha_3R_{\rho\lambda\sigma\delta}^2].
\end{equation}
Working to this same order, we may define an effective cosmological constant
\begin{equation}
g_{\rm eff}^2=g^2[1+\ft23(10\alpha_1+2\alpha_2+\alpha_3)g^2],
\label{eq:geff}
\end{equation}
so that
\begin{equation}
R_{\mu\nu}+4g_{\rm eff}^2g_{\mu\nu}=\alpha_3(-2C_{\mu\rho\lambda\sigma}
C_\nu{}^{\rho\lambda\sigma}+\ft13g_{\mu\nu}C_{\rho\lambda\sigma\delta}^2),
\label{eq:rsqlin}
\end{equation}
where we made the substitution $R_{\mu\nu\lambda\sigma}=C_{\mu\nu\lambda\sigma}
-g^2(g_{\mu\lambda}g_{\nu\sigma}-g_{\mu\sigma}g_{\nu\lambda})+\cdots$
which is a consequence of the zeroth order Einstein equation, $R_{\mu\nu}
=-4g^2g_{\mu\nu}+\cdots.$
We see that the coefficients $\alpha_1$ and $\alpha_2$ of $R^2$ and
$R_{\mu\nu}^2$, respectively, do not enter at linear order, so long as
we use the effective cosmological constant given by $g_{\rm eff}$.  This
is related to the fact that these two terms may be removed by a field
redefinition of the form $g_{\mu\nu}\to g_{\mu\nu}+a g_{\mu\nu}R+b R_{\mu\nu}$
with appropriate constants $a$ and $b$.

Although neutral black hole solutions may be obtained directly from
(\ref{eq:rsqlin}), we are mainly interested in $R$-charged solutions which
may be obtained from the full equations (\ref{eq:max}) and (\ref{eq:einslin}).
We turn to this in the next section.

\section{$R$-charged black holes}

The two-derivative Lagrangian, (\ref{eq:lag0}), admits a well-known
two-parameter family of static, stationary AdS$_5$ black hole solutions,
given by \cite{Behrndt:1998ns,Behrndt:1998jd}
\begin{eqnarray}
&&ds^2 = -H^{-2}fdt^2 + H(f^{-1}dr^2 + r^2d\Omega_{3}^{2}),\nonumber\\
&&A=\sqrt3\coth\beta\left(\fft1H-1\right)dt,
\label{eq:adsbh}
\end{eqnarray}
where the functions $H$ and $f$ are
\begin{eqnarray}
f&=&1-\fft\mu{r^2}+g^2r^2H^3,\nonumber\\
H&=&1 + \frac{\mu\sinh^2\beta}{r^2}.
\end{eqnarray}
The parameter $\mu$ is a non-extremality parameter, while $\beta$ is related
to the electric charge of the black hole.
The extremal (BPS) limit is obtained by taking $\mu\to\infty$ and $\beta\to0$ with $Q\equiv\mu\sinh^2\beta$ fixed, so that $f=1+g^2r^2H^3$ with $H=1+Q/r^2$.
These extremal solutions are naked singularities, and may be interpreted
as `superstars' \cite{Myers:2001aq}.  In the absence of higher-derivative
corrections, the BPS solutions may be smoothed out by turning on angular
momentum to form true black holes
\cite{Gutowski:2004ez,Gutowski:2004yv,Chong:2005hr,Kunduri:2006ek}

\subsection{The first order solution}

We wish to find the first order corrections to the $R$-charged black hole
solution given by (\ref{eq:adsbh}).  To do so, we treat the coefficients
($\alpha_1$, $\alpha_2$, $\ldots$, $\gamma_3$) of the four-derivative terms
in (\ref{eq:hdlag}) as small parameters, and make the ansatz
\begin{eqnarray}
&&ds^2 = -H^{-2}fdt^2 + H(f^{-1}dr^2 + r^2d\Omega_{3}^{2}),\nonumber\\
&&A=\sqrt3\coth\beta\left(\fft{1+a_1}H-1\right)dt,
\label{eq:ansz}
\end{eqnarray}
where
\begin{eqnarray}
H&=&1+\fft{\mu\sinh^2\beta}{r^2}+h_1,\nonumber\\
f&=&1-\fft{\mu}{r^2}+g^2r^2H^3+f_1.
\end{eqnarray}
Here, we treat $h_1$, $f_1$ and $a_1$ as small corrections, and will solve
for them to linear order in the parameters of the higher-derivative Lagrangian.
Note that this ansatz was designed so that the zeroth order equations are
automatically satisfied in the absence of $h_1$, $f_1$ and $a_1$.

Even after linearization in the small parameters, the individual equations of
motion, (\ref{eq:max}) and (\ref{eq:einslin}), yield complicated coupled
equations for the first order corrections.  However, the use of certain
symmetries of these equations yields tractable equations.  In particular,
the difference between the $tt$ and $rr$ components of the Einstein equation,
$R_t^t-R_r^r$, gives a second order equation involving only $h_1$, which is
easily solved.  The solution for $h_1$ can then be inserted into the Maxwell
equation, (\ref{eq:max}), to obtain a solution for $a_1$.  Finally, the
remaining components of the Einstein equation can be solved for $f_1$, thus
yielding the full solution.  The result is
\begin{eqnarray}
\label{eq:h1soln}
h_1&=&\fft{\mu^2\sinh^22\beta}{6H_0^2r^6}\bigl(7\alpha_1 + 5\alpha_2
+ 13\alpha_3 + 42\gamma_1 -12\gamma_2 + 12\gamma_3\bigr),\\
\label{eq:a1soln}
a_1&=&\fft{\mu^2\sinh^22\beta}{6H_0^3r^6}\biggl[\bigl(7\alpha_1 + 5\alpha_2
+13\alpha_3+42\gamma_1-12\gamma_2-12\gamma_3\tanh^2\beta\bigr)\nonumber\\
&&\kern6em +\fft{\mu\sinh^2\beta}{2r^2}\bigl(7\alpha_1+5\alpha_2+13\alpha_3
\nonumber\\
&&\kern11em+24(6\beta_1+3\beta_2+2\gamma_1-\gamma_2
+\gamma_3(1+\mathrm{sech}^2\beta))\bigr)\biggr],\qquad\\
\label{eq:f1soln}
f_1&=&\ft23g^4\bigl(10\alpha_1+2\alpha_2+\alpha_3\bigr)r^2H_0^3\nonumber\\
&&+\frac{g^2\mu^2\sinh^22\beta}{r^4}
\bigl(10\alpha_1-\alpha_2-13\alpha_3+20\gamma_1-\gamma_2-6\gamma_3\bigr)
\nonumber\\
&&+\fft{\mu^2}{r^6H_0}\left[\sinh^22\beta\bigl(3\alpha_1-\alpha_3
+18\gamma_1-3\gamma_2\bigr)+2\alpha_3\right]\nonumber\\
\nonumber\\
&&-\fft{\mu^3\sinh^22\beta\cosh^22\beta}{2r^8H_0^2}
\bigl(5\alpha_1+\alpha_2+\alpha_3+30\gamma_1-6\gamma_2\bigr)\nonumber\\
&&+\fft{\mu^4\sinh^42\beta}{96r^{10}H_0^3}
\bigl(47\alpha_1+13\alpha_2+17\alpha_3-144\beta_1-72\beta_2+276\gamma_1
-48\gamma_2-24\gamma_3\bigr),\nonumber\\
\end{eqnarray}
where $H_0=1+\mu\sinh^2\beta/r^2$ is the zeroth order solution for $H$.
(Since $h_1$, $a_1$ and $f_1$ are already linear in the parameters of the
higher order corrections, we may use $H$ and $H_0$ interchangeably in
the above expressions.)  Note that the first line in $f_1$ reproduces the
shift of the cosmological constant $g^2\to g_{\rm eff}^2$ given in
(\ref{eq:geff}).  This allows us to write
\begin{equation}
f=1-\fft\mu{r^2}+g_{\rm eff}^2r^2H^3+\bar f_1,
\label{eq:feff}
\end{equation}
where $\bar f_1$ is given by the remaining terms in (\ref{eq:f1soln}).

In obtaining the above solution, we have imposed the boundary conditions
that $h_1$ and $a_1$ both fall off faster than $1/r^2$ as $r\to\infty$ so
that the $R$-charge is not modified from its zeroth order value.  For
$f_1$, the boundary condition is taken as (\ref{eq:feff}), with $\bar f_1$
falling off faster than $1/r^2$.

\section{Field Redefinitions}

As given in (\ref{eq:hdlag}), we have parameterized the four-derivative terms
in the Lagrangian in terms of the eight coefficients ($\alpha_1$, $\alpha_2$,
$\ldots$, $\gamma_3$).  However, not all of these coefficients are physical.
This is because some of the terms in the higher derivative Lagrangian can be
removed by field redefinition.

To proceed, we consider transformations of the form
\begin{eqnarray}
g_{\mu\nu}&\to&g_{\mu\nu}+a(R+20g^2)g_{\mu\nu}+b(R_{\mu\nu}+4g^2g_{\mu\nu})
+cF_{\mu\lambda}F^\lambda{}_\nu + d F^2g_{\mu\nu},\nonumber\\
A_\mu&\to&(1+g^2(25a+5b-12c+60d)) A_\mu.
\label{eq:fredef}
\end{eqnarray}
Note that the first two terms in the metric shift incorporate the cosmological
constant; this corresponds to the zeroth order Einstein equation in the absence
of gauge excitations.  While this shift by the cosmological constant is not
strictly speaking necessary in performing the field redefinition, we
nevertheless find it convenient, as this avoids a shift in the effective
cosmological constant $g_{\rm eff}$ after the field redefinition.  In addition,
the scaling of the gauge field is chosen so that it will remain canonically
normalized after the shift of the metric.  The result of this transformation
is to shift the original Lagrangian (\ref{eq:lhd}) into
\begin{eqnarray}
e^{-1}\mathcal{L} &=& \left(1+12g^2(5a+b)\right)\biggl[R-\ft14F_{\mu\nu}^2
+ 12g^2\left(1-2g^2(5a+b)\right)\nonumber\\
&&\kern8em+\ft1{12\sqrt3}\left(1+3g^2(5a+b-12c+60d)\right)
\epsilon^{\mu\nu\rho\lambda\sigma}F_{\mu\nu}F_{\rho\lambda}A_\sigma\nonumber\\
&&\kern8em+\left(\alpha_1+\ft12(3a+b)\right)R^2+(\alpha_2-b)R_{\mu\nu}R^{\mu\nu}
+\alpha_3R_{\mu\nu\rho\sigma}R^{\mu\nu\rho\sigma}\nonumber\\
&&\kern8em+\left(\beta_1+\ft18(c-d)\right)(F_{\mu\nu}F^{\mu\nu})^2
+(\beta_2-\ft12c)
F^{\mu}{}_{\nu}F^{\nu}{}_{\rho}F^{\rho}{}_{\sigma}F^{\sigma}{}_{\mu}
\nonumber\\
&&\kern8em+\left(\gamma_1-\ft18(a+b+4c-12d)\right)RF^2\nonumber\\
&&\kern8em+\left(\gamma_2-\ft12(b+2c)\right)R_{\mu\nu}
F^{\mu\rho}F_{\rho}{}^{\nu} +
\gamma_3R_{\mu\nu\rho\sigma}F^{\mu\nu}F^{\rho\sigma}\biggr],
\end{eqnarray}
where, as usual, we only work to linear order in the shift parameters
($a$, $b$, $c$, $d$).

Up to an overall rescaling, this new Lagrangian can almost be brought back
to the original form, provided we shift the various coefficients as follows:
\begin{eqnarray}
&&g^2\to g^2\left(1+2g^2(5a+b)\right),\nonumber\\
&&\alpha_1\to\alpha_1-\ft12(3a+b),\kern6.1em\alpha_2\to\alpha_2+b,
\kern5.2em\alpha_3\to\alpha_3,\nonumber\\
&&\beta_1\to\beta_1-\ft18(c-d),\kern6.8em\beta_2\to\beta_2+\ft12c,\nonumber\\
&&\gamma_1\to\gamma_1+\ft18(a+b+4c-12d),\qquad\gamma_2\to\gamma_2+\ft12(b+2c),\qquad\gamma_3\to\gamma_3.
\label{eq:frtrans}
\end{eqnarray}
One difference remains, however, and that is the coefficient of the
$F\wedge F\wedge A$ Chern-Simons term.  This suggests that, when considering
higher derivative corrections in gauged supergravity, there is in fact a
preferred field redefinition frame where this Chern-Simons term remains
uncorrected.  (Such a preferred frame also shows up when considering the
supersymmetric completion of the mixed $\mbox{Tr}\,R\wedge R\wedge A$ term
\cite{Hanaki:2006pj}.)  This $F\wedge F\wedge A$ term is unimportant, however,
for the spherically symmetric $R$-charged black holes considered above in
Section~3.

Ignoring the $F\wedge F\wedge A$ term, the freedom to perform field
redefinitions of the form (\ref{eq:fredef}) indicates that at most four
of the eight coefficients of the higher derivative terms will be physical.
Clearly $\alpha_3$ and $\gamma_3$ are physical, as they cannot be removed
by the transformation of (\ref{eq:frtrans}).  The additional two physical
coefficients can be taken to be a linear combination of
\begin{equation}
\hat\beta_1\equiv\beta_1+\ft1{144}(\alpha_1-7\alpha_2)
+\ft1{12}(\gamma_1+\gamma_2)
\qquad\hbox{and}\qquad
\hat\beta_2\equiv\beta_2+\ft14\alpha_2-\ft12\gamma_2.
\label{eq:hb1b2}
\end{equation}
In addition, although $g^2$ is shifted by the field redefinition, the physical
cosmological constant, $g^2_{\rm eff}$, as defined in (\ref{eq:geff}),
remains invariant.

The use of field redefinitions allows us to rewrite the four-derivative
Lagrangian in various forms.  A common choice would be to use the Gauss-Bonnet
combination $R^2-4R_{\mu\nu}^2+R_{\mu\nu\lambda\sigma}^2$ for the
curvature-squared terms.  This system has been extensively studied in
the absence of higher-derivative gauge field corrections, and has the feature
that it admits {\it exact} spherically symmetric black hole solutions, both
without \cite{Boulware:1985wk,Wheeler:1985nh} and with \cite{Wiltshire:1985us}
$R$-charge.  An alternate choice, which is perhaps more natural
from a supersymmetric point of view \cite{Hanaki:2006pj}, would be to use the
Weyl-squared combination $C_{\mu\nu\lambda\sigma}^2=\fft16R^2
-\fft43R_{\mu\nu}^2+R_{\mu\nu\lambda\sigma}^2$.  Either one of these choices
would fix two of the coefficients ({\it i.e.}~$\alpha_1$ and $\alpha_2$ in
terms of $\alpha_3$).  The additional freedom to perform field redefinitions
may then be used to eliminate the mixed $RF^2$ and $R_{\mu\nu}F^{\mu\lambda}
F^\lambda{}^\nu$ terms parameterized by $\gamma_1$ and $\gamma_2$.

\subsection{Field redefinitions and the first order solution}

Given the above field redefinition, it is instructive to examine its effect
on the first order black hole solution of (\ref{eq:h1soln}), (\ref{eq:a1soln})
and (\ref{eq:f1soln}).  In this case, it is straightforward to see that
the coefficient shift of (\ref{eq:frtrans}) results in
\begin{eqnarray}
h_1&\to&\tilde h_1=h_1+\fft{\mu^2\sinh^22\beta}{8H_0^2r^6}
(-7a+b+12c-84d),\nonumber\\
a_1&\to&\tilde a_1=a_1+\fft{\mu^2\sinh^22\beta}{8H_0^3r^6}\left[
(-7a+b+12c-84d)-\fft{3\mu\sinh^2\beta}{r^2}(a+b-4c+12d)\right],\nonumber\\
f_1&\to&\tilde f_1=f_1-2g^4(5a+b)r^2H_0^3-\fft{g^2\mu^2\sinh^22\beta}{2r^4}
(25a+8b-18c+60d)\nonumber\\
&&\kern4em
+\fft{3\mu^2\sinh^22\beta}{8r^6H_0}\biggl[-2(3a+b-8c+36d)
+\fft{\mu\cosh^22\beta}{r^2H_0}(5a+b-12c+60d)\nonumber\\
&&\kern11em-\fft{\mu^2\sinh^22\beta}{r^4H_0^2}(a-2c+12d)\biggr].
\label{eq:cssol}
\end{eqnarray}
At first, this result may appear somewhat surprising.  After all, this
field redefinition is supposed to be `unphysical', and yet the form of the
solution has changed.  The resolution of this puzzle lies in the fact that
the we have shifted the metric by terms that are not necessarily proportional
to the lowest order equations of motion.  (While we have taken care to
incorporate the cosmological constant in (\ref{eq:fredef}), we have omitted
the gauge field stress tensor in the shift.)  In this sense, while the
original and shifted metrics both solve the equations of motion, they
nevertheless correspond to physically distinct solutions.  The field
redefinition of (\ref{eq:fredef}) is then more naturally thought of as
a mapping between solutions.

More explicitly, we note that the shift of the metric given in
(\ref{eq:fredef}) takes the black hole solution away from the form of
the initial ansatz given by (\ref{eq:ansz}).  In particular, shifting
the metric by (\ref{eq:fredef}) and using the zeroth order
solution gives
\begin{eqnarray}
g_{tt}&\to&\tilde g_{tt}=g_{tt}\left[1 - \fft{\mu^2\sinh^22\beta}{2r^6H_0^3}
(a + 2b -6c+12d)\right],\nonumber\\
g_{rr}&\to&\tilde g_{rr}=g_{rr}\left[1 - \fft{\mu^2\sinh^22\beta}{2r^6H_0^3}
(a + 2b -6c+12d)\right],\nonumber\\
g_{\alpha\beta}&\to&\tilde g_{\alpha\beta}=
g_{\alpha\beta}\left[1-\fft{\mu^2\sinh^22\beta}{2r^6H_0^3}
(a-b+12d)\right],
\end{eqnarray}
where $\alpha$ and $\beta$ refer to coordinates on $S^3$.  It is now
possible to see that a coordinate transformation $r\to\tilde r$ is
necessary in order to restore the canonical form of the shifted metric.
By identifying
\begin{eqnarray}
d\tilde s^2 &=& \tilde{g}_{tt}dt^2 +
\tilde{g}_{rr}dr^2 +
\tilde{g}_{\theta\theta}d\Omega_3^2\nonumber\\
&=&-\tilde H^{-2}\tilde fdt^2+\tilde H(\tilde f^{-1}d\tilde r^2+\tilde r^2
d\Omega_3^2),
\end{eqnarray}
we end up with expressions for $\tilde H$ and $\tilde f$
\begin{equation}
\tilde H=\fft{\tilde g_{\theta\theta}}{\tilde r^2},\qquad
\tilde f=-\tilde g_{tt}\tilde g_{\theta\theta}^2{\tilde r^4},
\label{eq:hftmet}
\end{equation}
as well as a differential equation relating $\tilde r^2$ with $r^2$
\begin{equation}
\fft{d(\tilde r^2)}{d(r^2)}=\fft{\tilde g_{tt}\tilde g_{rr}
\tilde g_{\theta\theta}}{r^2}.
\end{equation}
Note that, in defining the angular coordinate $\theta$, we have taken the
metric on the unit $S^3$ to be of the form $d\Omega_3^2=d\theta^2+\cdots.$
The equation for $\tilde r^2$ is easily solved, and yields the relation
\begin{equation}
\tilde{r}^2 = r^2\left[1 +
\fft{3\mu^2\sinh^22\beta}{8r^6H_0^2}(3a +3b-12c+36d)\right],
\label{eq:rtrel}
\end{equation}
where we have set a possible integration constant to zero to preserve the
$r\to\infty$ asymptotics.

We are now able to explicitly compute the shifted metric functions $\tilde h_1$
and $\tilde f_1$ as well as the shifted gauge potential $\tilde a_1$.  For
$\tilde h_1$, we use the definition
\begin{equation}
\tilde H=1+\fft{\mu\sinh^2\beta}{\tilde r^2}+\tilde h_1,
\end{equation}
along with (\ref{eq:hftmet}) and (\ref{eq:rtrel}) to obtain
\begin{equation}
\tilde h_1=h_1+\fft{\mu^2\sinh^22\beta}{8H_0^2r^6}(-7a+b+12c-84d),
\end{equation}
which is in perfect agreement with (\ref{eq:cssol}).  For $\tilde f_1$, on
the other hand, we find
\begin{eqnarray}
\tilde f_1&=&f_1-2g^4(5a+b)r^2H_0^3
-\fft{3g^2\mu^2\sinh^22\beta}{2r^4}(b-2c)\nonumber\\
&&\qquad+\fft{3\mu^2\sinh^22\beta}{8r^6H_0}\biggl[
-2(3a+b-8c+36d)+\fft{\mu\cosh2\beta}{r^2H_0}(5a+b-12c+60d)\nonumber\\
&&\kern9em-\fft{\mu^2\sinh^22\beta}{r^4H_0^2}(a-2c+12d)\biggr].
\label{eq:ftcomp}
\end{eqnarray}
Note that we have defined $\tilde f_1$ by
\begin{equation}
\tilde f=1-\fft\mu{\tilde r^2}+\tilde g^2\tilde r^2\tilde H^3+\tilde f_1,
\label{eq:ftdef1}
\end{equation}
where $\tilde g^2=g^2(1+2g^2(5a+b))$ is the shifted cosmological constant
given in (\ref{eq:frtrans}).

Comparison of (\ref{eq:ftcomp}) with (\ref{eq:cssol}) clearly demonstrates
a difference in the $\mathcal O(g^2)$ term.  The origin of this difference
is somewhat subtle, and is related to the choice of boundary conditions for
the shifted and unshifted solutions.  To see this, we recall that the
gauge potential $A_\mu$ is also shifted by the field redefinition
(\ref{eq:fredef}) so that it maintains canonical normalization.  The
implication of this shift on the black hole solution is that
\begin{equation}
A_t\to \bigl(1+g^2(25a+5b-12c+60d)\bigr)A_t,
\end{equation}
where
\begin{equation}
A_t=\sqrt3\coth\beta\left(\fft{1+a_1}H-1\right),\qquad
H=1+\fft{\mu\sinh^2\beta}{r^2}+h_1.
\end{equation}
In order to rescale the potential without adding any $\mathcal O(1/r^2)$
terms to $H_0$, $h_1$ or $a_1$, we must instead shift the two parameters
$\mu$ and $\beta$ of the black hole according to
\begin{equation}
\coth\beta\to\coth\beta\bigl(1+g^2(25a+5b-12c+60d)\bigr),\qquad
\mu\sinh^2\beta\to\mu\sinh^2\beta.
\label{eq:mubres}
\end{equation}
This corresponds to a rescaling of the nonextremality parameter $\mu$
\begin{equation}
\mu\to\tilde\mu=\mu(1+2g^2\cosh^2\beta(25a+5b-12c+60d)).
\end{equation}
In this case, the shifted metric function $\tilde f$, given in
(\ref{eq:ftdef1}), ought to more properly be written as
\begin{equation}
\tilde f=1-\fft{\tilde\mu}{\tilde r^2}+\tilde g^2\tilde r^2\tilde H^3
+\hat f_1,
\end{equation}
where
\begin{eqnarray}
\hat f_1&=&\tilde f_1+\fft{2g^2\mu\cosh^2\beta}{r^2}(25a+5b-12c+60d)
\nonumber\\
&=&f_1+\lambda\fft{H_0}{r^2}-2g^4(5a+b)r^2H_0^3
-\fft{g^2\mu^2\sinh^22\beta}{2r^4}(25a+8b-18c+60d)+\cdots.\nonumber\\
\end{eqnarray}
This now agrees with $\tilde f_1$ of (\ref{eq:cssol}) up to a solution
$\lambda H_0/r^2$ to the homogeneous differential equation for $f_1$, where
\begin{equation}
\lambda=2g^2\mu\cosh^2\beta(25a+5b-12c+60d).
\end{equation}
This is a modification of the $\mathcal O(1/r^2)$ term in $f_1$, which,
however, is subdominant in $f$, as the leading behavior of $f$ is given by
$f\sim g^2_{\rm eff}r^2$ for an asymptotically Anti-de Sitter background.

Finally, we may follow the effect of the field redefinition (\ref{eq:fredef})
on the gauge potential term $a_1$.  Given the $\mu$ and $\beta$ rescaling of
(\ref{eq:mubres}), we obtain 
\begin{equation}
\tilde a_1=(1+a_1)\fft{\tilde H}H-1.
\end{equation}
Working out the right hand side of this expression, we find that it
agrees with (\ref{eq:cssol}).  We have thus seen that the first
order solution for the spherically symmetric $R$-charged black hole indeed
transforms as expected under field redefinitions.

\section{Discussion}

While we have considered general field redefinitions given by four
parameters ($a$, $b$, $c$, $d$), a preferred subset of this would be to
shift the metric by the full zeroth order equation of motion
\begin{equation}
R_{\mu\nu}+4g^2g_{\mu\nu}-\ft12F_{\mu\lambda}F_\nu{}^\lambda
+\ft1{12}g_{\mu\nu}F^2.
\end{equation}
In the above notation, this corresponds to taking
\begin{equation}
c=\ft12b,\qquad d=-\ft1{12}(a-b).
\end{equation}
In this case, we may redefine the coefficients of the higher derivative
terms according to
\begin{eqnarray}
\beta_1&=&\hat\beta_1-\ft1{12}(\hat\gamma_1+\hat\gamma_2)
+\ft1{144}(\alpha_1-7\alpha_2),\nonumber\\
\beta_2&=&\hat\beta_2+\ft12\hat\gamma_2+\ft14\alpha_2,\nonumber\\
\gamma_1&=&\hat\gamma_1-\ft16(\alpha_1-\alpha_2),\nonumber\\
\gamma_2&=&\hat\gamma_2+\alpha_2,
\end{eqnarray}
so that the set ($\alpha_3$, $\hat\beta_1$, $\hat\beta_2$, $\hat\gamma_1$,
$\hat\gamma_2$, $\gamma_3$) are invariant under the restricted field
redefinitions.  Note that $\hat\beta_1$ and $\hat\beta_2$ are the physical
coefficients previously defined in (\ref{eq:hb1b2}).

It is illuminating to rewrite the higher derivative Lagrangian (\ref{eq:lhd})
in terms of the new parameters.  Ignoring the Chern-Simons term, the result is
\begin{eqnarray}
\label{eq:plag}
e^{-1}\mathcal L&=&\bigl(1-8g^2(5\alpha_1+\alpha_2)\bigr)\biggl[
R-\ft14\hat F^2+12g_{\rm eff}^2+\alpha_1\mathcal E^2
+\alpha_2\mathcal E_{\mu\nu}^2
+\alpha_3\bigl(R_{\mu\nu\lambda\sigma}^2-8g^4\bigr)\nonumber\\
&&\kern6em+\hat\beta_1(\hat F^2)^2+\hat\beta_2\hat F^4
+\hat\gamma_1\mathcal E\hat F^2-\hat\gamma_2\mathcal E_{\mu\nu}
\hat F^{\mu\sigma}\hat F^\nu{}_\sigma+\gamma_3R^{\mu\nu\lambda\sigma}
\hat F_{\mu\nu}\hat F_{\lambda\sigma}\biggr],\nonumber\\
\end{eqnarray}
where
\begin{equation}
\mathcal E_{\mu\nu}\equiv R_{\mu\nu}+4g^2g_{\mu\nu}
-\ft12\hat F_{\mu\lambda}\hat F_\nu{}^\lambda+\ft1{12}g_{\mu\nu}\hat F^2,\qquad
\mathcal E=\mathcal E^\mu{}_\mu
\end{equation}
is the zeroth order equation of motion.  Note that we have worked to linear
order in pulling out the overall factor $1-8g^2(5\alpha_1+\alpha_2)$
renormalizing Newton's constant.  Furthermore, $\hat F=d\hat A$ is
a rescaled field strength defined by
\begin{equation}
\hat A_\mu=\left[1+8g^2\bigl(\ft13(5\alpha_1+\alpha_2)+5\hat\gamma_1
-\hat\gamma_2\bigr)\right]A_\mu,
\end{equation}
so that $\hat A_\mu$ remains invariant under the field redefinition of
(\ref{eq:fredef}).  The structure of (\ref{eq:plag}) now clearly demonstrates
that, of the four-derivative terms, only those parameterized by
($\alpha_3$, $\hat\beta_1$, $\hat\beta_2$, $\gamma_3$) are physical, as the
remaining terms are manifestly proportional to the zeroth order equation of
motion.

In principle, the choice of field redefinitions allows us to go back and forth
between the Gauss-Bonnet and Weyl-squared parameterizations of the
higher-derivative terms in the Lagrangian.  In this sense, it is perhaps not
a complete surprise to see that in some cases both parameterizations yield
the same results for the entropy of BPS black holes
\cite{Guica:2005ig,Castro:2007hc,Cvitan:2007en}, even though the bare
Gauss-Bonnet correction is not supersymmetric in itself.  (Of course, the
bare Weyl-squared term is not supersymmetric by itself either.)  What this
suggests is that the Riemann-squared term parameterized by $\alpha_3$ plays
a crucial and perhaps dominant role in the geometry of higher-derivative
black holes, and that the additional matter and auxiliary field terms may
contribute only indirectly through their effects on the geometry, at least
in the BPS case where there is additional symmetry at the horizon.

Finally, given the general higher-derivative corrected $R$-charged black holes,
it would be interesting to study their thermodynamics and hydrodynamics.  One
outcome of this study ought to be a clear identification of physical versus
unphysical parameters of the theory.  In particular, in the parameterization of
(\ref{eq:plag}), we would expect all dependence on ($\alpha_1$, $\alpha_2$,
$\hat\gamma_1$, $\hat\gamma_2$) to drop out of the thermodynamical quantities.
One difficulty in exploring the higher-derivative theory is that some care
must be taken in generalizing the Gibbons-Hawking surface term (which we have
ignored throughout this letter).  This is because the general ({\it i.e.}~non
Gauss-Bonnet) combination of $R^2$ terms leads to higher than second-derivative
terms in the equations of motion, and hence necessitates specifying additional
boundary data \cite{Myers:1987yn}.  As demonstrated in \cite{Buchel:2004di},
one way around this is to perturb in the higher-derivative terms and to
demand that the undesired boundary variations vanish when the lowest-order
equations of motion are imposed.  We are currently applying this procedure
to the general parameterized four-derivative Lagrangian with a goal of
exploring higher-derivative black hole thermodynamics using holographic
renormalization.

\section*{Acknowledgments}

We wish to thank A. Castro, J. Davis and K. Hanaki for useful conversations.
This work was supported in part by the US Department of Energy under
grant DE-FG02-95ER40899.


\end{document}